\documentclass[twocolumn, aps, floatfix]{revtex4-1}
\usepackage[latin2]{inputenc}
\usepackage{amsfonts}
\usepackage{amsmath}
\usepackage{amssymb}
\usepackage{amsthm}
\usepackage[T1]{fontenc}
\usepackage[dvips]{graphicx}
\date{2006-02-13}
\begin{document}

\title{Synthetic magnetic fluxes on the honeycomb lattice}
\author{Agnieszka G\'orecka,$^1$ Beno\^{\i}t Gr\'{e}maud,$^{1,2,3}$ and Christian Miniatura$^{1,2,4}$
}                      

\affiliation{
\mbox{$^1$ Centre for Quantum Technologies, National University of Singapore, 3 Science Drive 2, Singapore 117543, Singapore} \\
\mbox{$^2$ Department of Physics, National University of Singapore, 2 Science Drive 3, Singapore 117542, Singapore} \\
\mbox{$^3$ Laboratoire Kastler Brossel, Ecole Normale Sup\'{e}rieure, CNRS, UPMC; 4 Place Jussieu, 75005 Paris, France} \\
\mbox{$^4$ Institut Non Lin\'{e}aire de Nice, UMR 6618, UNS, CNRS; 1361 route des Lucioles, 06560 Valbonne, France} \\
}             

\date{\today}

\begin{abstract}
We devise experimental schemes able to mimic uniform and staggered magnetic fluxes acting on ultracold two-electron atoms, such as ytterbium atoms, propagating in a honeycomb lattice. The atoms are first trapped into two independent state-selective triangular lattices and are further exposed to a suitable configuration of resonant Raman laser beams. These beams induce hops between the two triangular lattices and make atoms move in a honeycomb lattice. Atoms traveling around each unit cell of this honeycomb lattice pick up a nonzero phase. In the uniform case, the artificial magnetic flux sustained by each cell can reach about two flux quanta, thereby realizing a cold atom analogue of the Harper model with its notorious Hofstadter's butterfly structure. Different condensed-matter phenomena such as the relativistic integer and fractional quantum Hall effects, as observed in graphene samples, could be targeted with this scheme.
\end{abstract}


\maketitle

\section{introduction}
\label{in}

The ability to produce and trap dilute degenerate Bose and Fermi gases \cite{ketterle1, ketterle2} ushered ultracold atoms as powerful players to explore phenomena mostly studied until now in condensed-matter physics, see \cite{lewenstein, bloch1} for comprehensive reviews. Paradigmatic examples are the observation of the Abrikosov vortex lattice \cite{madison, abo}, the Mott-superfluid transition \cite{greiner1}, the BEC-BCS crossover \cite{greiner2, bourdel} and the Kosterlitz-Thouless transition \cite{hadzibabic}.

One current motive in the field is to address the physics of two-dimensional (2D) electrons exposed to a magnetic field and the associated integer (IQHE) and fractional (FQHE) quantum Hall effects \cite{prange}. Even though atoms are neutral, people have rapidly realized that, because of the similarity between the Lorentz and the Coriolis forces, repulsively interacting atoms loaded in rotating traps could target the fractional Hall effect regime \cite{cooper}. Unfortunately the first attempts failed as it is difficult to reach the regime of one vortex per particle \cite{schweikhard, bretin}. One key reason is that the observation of the Laughlin states requires a very fine tuning of the rotation frequency compared to the trap frequency.

In the meantime other promising methods using light-induced gauge fields \cite{zoller, ruseckas, juzeliunas, spielman, dalibard} where theoretically explored, see \cite{dalibard1} for a review. For atoms trapped in 2D optical lattices, these schemes amount to control the phases of the hopping amplitudes such that non-intersecting loops acquire a nonzero phase. This simply mimic the effect of a constant magnetic field leading to Harper's model \cite {harper} and Hofstadter's butterfly \cite{hofstadter}. More elaborate schemes targeting non-Abelian gauge fields have been proposed \cite{osterloh, bermudez}. From an experimental point of view, the successful generation of an Abelian gauge field in the bulk has been reported with the observation of vortices \cite{lin1, lin2}.

The aim of our work is to devise workable experimental schemes that would create artificial Abelian gauge fields acting on a honeycomb lattice filled with earth-alkaline atoms. These artificial gauge fields mimic effective magnetic fields imposing uniform or staggered fluxes through the lattice. Related schemes have been theoretically proposed in \cite{dalibard, lim} for the rectangular lattice. Recently a scheme mimicking an effective periodic magnetic field has been proposed in~\cite{cooper1}. Our primary interest in the honeycomb lattice is that its ground and first-excited bands give rise to two Dirac points around which the dispersion is linear. When loaded with ultracold fermions around half-filling, the Fermi energy slices the band structure around the Dirac points and one gets a cold atom analogue of graphene \cite{zhu, keanloon}. Key experiments with graphene samples have revealed a particular quantum Hall effect \cite{novoselov, zhang}. This is due to the relativistic nature of the low-energy electronic excitations which behave like massless Dirac particles \cite{goerbig}. Recently the first observations of a FQHE state at $\nu = \pm 1/3$ in suspended graphene were reported \cite{bolotin, du}. Because of the richness and flexibility of the cold atoms technology, we believe that experiments where fermions, bosons or fermion-boson mixtures are loaded in the honeycomb optical lattice and are exposed to an artificial magnetic field yield situations difficult to explore in graphene research, in particular when the hopping amplitudes are imbalanced \cite{guangquan}. It is worth mentioning that configurations giving rise to Dirac points in the square lattice have been proposed~\cite{hou, goldman}. However it can be shown that the Dirac points are robust in the honeycomb optical lattice where the laser beams need not be at perfect respective in-plane $120^\circ$ angles from each other or carrying exactly the same intensities~\cite{keanloon}.  In this context, it is worth noticing that the first experimental realization of spin-state dependent lattices realizing a honeycomb lattice and loaded with ultracold $^{87}$Rb ground-state atoms has been recently reported~\cite{soltan}.

The paper is organized as follows. In Section~\ref{lattice} we introduce the laser configuration producing the two uncoupled state-dependent triangular lattices we need.
In Section~\ref{uni}, we introduce the Raman laser configuration which induce hops between the two previous sublattices, thereby realizing a (slightly non regular) honeycomb lattice where the atoms can move. We show that a single Raman laser scheme is not sufficient to induce an artificial magnetic field and we propose a 4-beam scheme to achieve a uniform flux through the lattice. About two flux quanta per unit cell can be generated with our scheme, thereby reaching the strong field limit and realizing the cold atom analogue of the Harper model. In Section~\ref{stg} we introduce two set-ups that allow to achieve a staggered flux through the lattice similar to the one shown in \cite{dalibard}. About one flux quanta per unit cell can be generated with both schemes. In Section~\ref{harpermodel}, we give the associated Harper models. We summarize and conclude in Section~\ref{sum}.

\section{State-dependent triangular optical lattices}
\label{lattice}

In this section we introduce the laser configuration that produces two triangular state-dependent optical lattices. Like in~\cite{dalibard}, all our calculations are done for the bosonic isotope $^{174}$Yb of ytterbium \cite{barber, takasu}. We restrict our analysis to two internal states, namely the ground state $^{1}\!S_{0}$, hereafter denoted by $|g\rangle$, and the long-lived (lifetime about 20 s) metastable excited state $^{3}\!P_{0}$, hereafter denoted by $|e\rangle$. The energy separation between these two states is $\hbar\omega_0=\hbar c k_0 = hc/\lambda_0$, where $\lambda_0=0.578\mu$m and the so-called "magic" and "anti-magic" wavelengths of this isotope are $\lambda_m=0.76 \mu$m and $\lambda_{am}=1.12 \mu$m. 

In a nutshell, our strategy is the following. A first laser configuration working at the "magic" frequency creates a honeycomb potential with lattice constant $a\propto \lambda_m$ which confines both Yb internal states in its minima. At this stage there is thus no special spatial organization of the Yb internal states among the two triangular sublattices of the honeycomb lattice. A second laser configuration working at the "anti-magic" frequency is then superimposed to the previous one to create an optical standing-wave potential.
When $3a=\lambda_{am}$, as a net result of the combination of these two optical potentials, state-$|g\rangle$ atoms are trapped in the minima of a triangular lattice while state-$|e\rangle$ atoms are trapped in the minima of another shifted triangular lattice. As a whole, we get a (slightly non regular) honeycomb lattice where state-$|g\rangle$ atoms are solely trapped in one of its sublattice while the state-$|e\rangle$ atoms are trapped in the other. Considering the hexagonal Bravais Wigner-Seitz cell of the new honeycomb lattice, this means that its vertices are now alternatively occupied by state-$|g\rangle$ and state-$|e\rangle$ atoms. At this stage however, the state-$|g\rangle$ and state-$|e\rangle$ sublattices are still uncoupled as atoms trapped in one sublattice cannot flip their internal state and hop on the other sublattice. How to couple these two sublattices will be the topic of Section~\ref{uni}.

Throughout this Section and Section~\ref{uni}, for convenience purposes, we will use $\lambda_{am}/2 = 0.56 \mu m$ as our space unit.

\subsection{"Magic" honeycomb lattice}

\begin{figure}[!htb]
\centerline{\includegraphics[width=7cm]{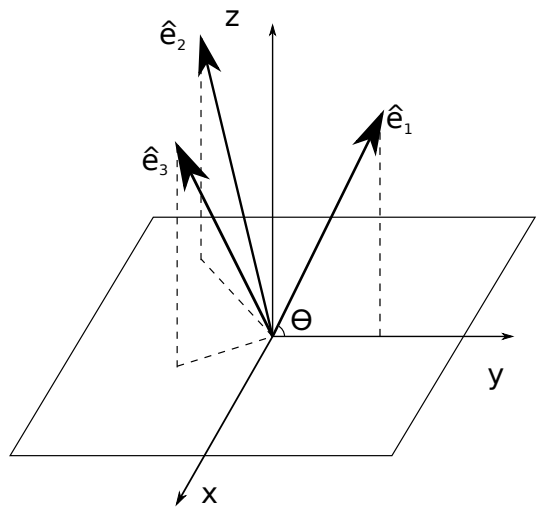}}
\caption{\label{f1} 
The laser configuration creating a honeycomb optical lattice in the ($Ox,Oy$) plane at the magic wavelength. It is made of three monochromatic linearly-polarized beams with the same off-plane elevation angle $\theta$. The in-plane projections of the laser wavevectors have respective angles $120^\circ$.
}
\end{figure}

The basic laser configuration creating the honeycomb lattice was presented and analyzed in \cite{keanloon}. In the present case, we choose a slightly modified version of it by considering the superposition of three linearly-polarized running monochromatic waves at angular frequency $\omega_m = ck_m = 2\pi c/\lambda_m$ with wavevectors ${\bf k}_i = k_m {\bf \hat{e}}_i$, where:

\begin{subequations}
\label{vectorsk}
\begin{align}
&{\bf \hat{e}}_1 = \cos\theta \, {\bf \hat{e}}_y + \sin\theta \, {\bf \hat{e}}_z,\\
&{\bf \hat{e}}_2 = - \cos\theta \frac{\sqrt{3} \bf \hat{e}_x + {\bf \hat{e}}_y}{2}+ \sin\theta \, {\bf \hat{e}}_z,\\
&{\bf \hat{e}}_3 = \cos\theta \frac{\sqrt{3} \bf \hat{e}_x - {\bf \hat{e}}_y}{2}+ \sin\theta \, {\bf \hat{e}}_z,
\end{align}
\end{subequations}
$\theta$ being the elevation angle of the "magic" beams off the $(Ox,Oy)$ plane, see Fig.~\ref{f1}.

Up to an inessential additive constant, the resulting optical dipole potential $V_m({\bf r})$ is translation-invariant along the $Oz$ direction and displays a honeycomb structure in the $(Ox,Oy)$ plane with lattice constant $a=2\lambda_m/(3\sqrt{3}\cos\theta)$. After a suitable choice of space and time origins, it is then given by $V_m({\bf r})= V_0 [3+2 f_m({\bf r})]$ where:
\begin{equation}
f_m({\bf r})=\cos{({\bf b}_1\cdot {\bf r})}\!+\!\cos{({\bf b}_2\cdot {\bf r})}\!+\!\cos{(({\bf b}_1\!+\!{\bf b}_2)\cdot {\bf r})}),
\end{equation}
and where $V_0$ denotes the strength of the potential. The vectors ${\bf b}_1={\bf k}_3 - {\bf k}_1$ and ${\bf b}_2 = {\bf k}_1 - {\bf k}_2$ feature the primitive reciprocal lattice vectors of the honeycomb lattice. In turn they define the primitive honeycomb Bravais lattice vectors ${\bf a}_1$ and ${\bf a}_2$ through ${\bf a}_i\cdot{\bf b}_j=2\pi \delta_{ij}$. In terms of our space units, we find:
\begin{equation}
\label{eq:honeypot}
f_m({\bf r})=2\cos{(\pi \gamma x)} \, \cos{(\sqrt{3}\pi \gamma y)}+\cos{(2\pi \gamma x}),
\end{equation}
where:
\begin{equation}
\gamma = \frac{\sqrt{3} \lambda_{am}\cos\theta}{2\lambda_m}
\end{equation}

\begin{figure}[!htb]
\centerline{\includegraphics[width=6cm]{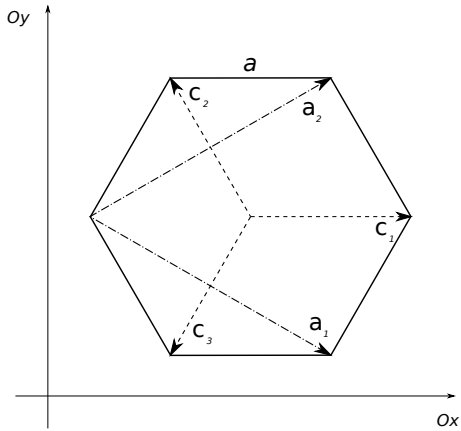}}
\caption{\label{f2} 
Hexagonal Wigner-Seitz cell of the honeycomb lattice associated to the optical potential generated by Eq. \eqref{eq:honeypot}. The figure shows the two primitive Bravais vectors ${\bf a_i}$, the three vectors ${\bf c_j}$ connecting each honeycomb lattice site to its three nearest neighbors and the honeycomb lattice constant $a$.
}
\end{figure}

Throughout this paper, we assume that a suitable external confinement along axis $Oz$ restricts the atomic dynamics in the ($Ox,Oy$) plane. Then, for $V_0>0$, the atoms are trapped in the minima of the potential which coincide with its zeroes. These are organized in a regular honeycomb structure made of two shifted identical triangular sublattices \cite{keanloon}. Each lattice site is connected to its three nearest neighbors by the vectors ${\bf c}_a$ ($a=1,2,3$) satisfying:
\begin{subequations}
\begin{align}
&{\bf a}_1 = {\bf c}_1-{\bf c}_2, \\
&{\bf a}_2 = {\bf c}_1-{\bf c}_3, \\
&\sum_a {\bf c}_a = {\bf 0}.
\end{align}
\end{subequations}
Fig.\ref{f2} shows the hexagonal Wigner-Seitz cell of this regular honeycomb lattice together with the primitive Bravais and nearest-neighbor vectors. The potential is maximum at its center, vanishes at each of its vertices and exhibits a saddle-point at the mid-point of each of its sides \cite{keanloon}.

\subsection{Creating state-dependent triangular sublattices}

The previous "magic" honeycomb potential trap atoms in its minima irrespective of their internal state. Starting from this situation, we would like now to selectively trap atoms with a given internal state in a given triangular sublattice of a new honeycomb lattice. As a result the six vertices of the corresponding Wigner-Seitz cell would alternate trapped internal states. This is achieved by shining two counter-propagating laser beams along the $Ox$ direction, their common angular frequency being $\omega_{am} = ck_{am} = 2\pi c/\lambda_{am}$. Taking the origin of space at a point where the "magic" honeycomb potential $V_m({\bf r})$ is maximum, this "anti-magic" standing-wave potential is given by:
\begin{equation}
V_{am}({\bf r})=V_1 \frac{1+\cos{({\bf b}\cdot({\bf r}-{\bf r}_0))}}{2}
\end{equation}
where ${\bf b}=2k_{am} \bf \hat{e}_x$ is the anti-magic reciprocal lattice vector, $V_1$ being the corresponding potential strength. The position ${\bf r}_0$ where this potential reaches its maximum value is determined by the relative phase of the two interfering "anti-magic" laser beams. The effective potentials for the ground and excited states then read respectively:
\begin{subequations}
\begin{align}
&V_g({\bf r})=V_m({\bf r})+V_{am}({\bf r}),\\
&V_e({\bf r})=V_m({\bf r})-V_{am}({\bf r}).
\end{align}
\end{subequations}
The strategy is now to find parameters for which both $V_g$ and $V_e$ sustain the same (regular) triangular Bravais lattice, both potential minima being organized in two shifted triangular lattices. As a whole, one would get a honeycomb lattice where adjacent sites would now be loaded with atoms having different internal states, or, in other words, where each sublattice sustains a given internal state. These sublattices of minima will be triangular if the "magic" and "anti-magic" lattices match. This the case if ${\bf b}$ belongs to the reciprocal lattice of the "magic" honeycomb lattice. This requirement is simply met when ${\bf b}={\bf b}_1+{\bf b}_2$, i.e. when $\lambda_{am}=3a$ (or equivalently $\gamma = 1$). This is implemented by stretching the honeycomb lattice and choosing the common polar angle $\theta$ of the "magic" laser beams to satisfy:
 \begin{equation}
\cos\theta = \frac{2\lambda_m}{\sqrt{3}\lambda_{am}}.
\end{equation}
For the chosen internal states of the Yb atom, one finds the required off-plane angle $\theta=31.41^o$.

\begin{figure}[!htb]
\centerline{\includegraphics[width=7.5cm]{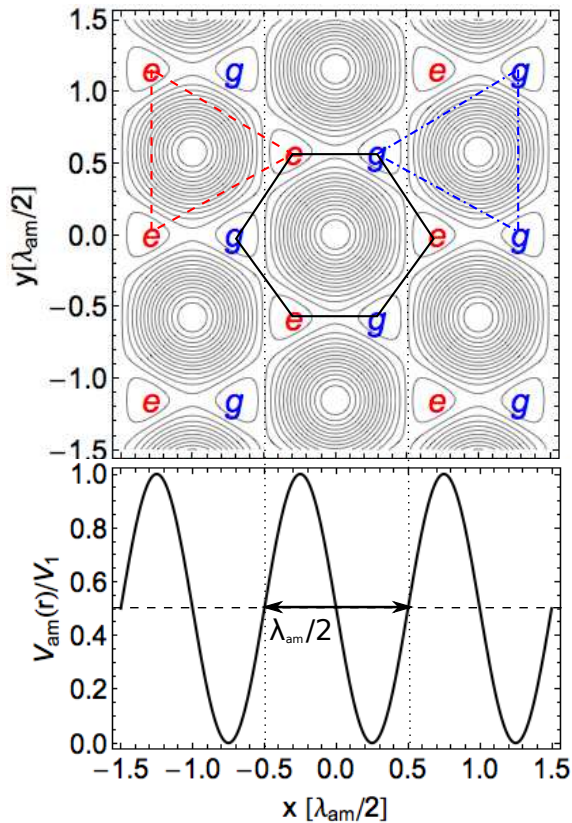}}
\caption{\label{f3} [Color online] Top: contour plot of the stretched "magic" honeycomb potential $V_m$ ($a=2/3$ in space units of $\lambda_{am}/2$). Bottom: plot of the "anti-magic" standing-wave potential $V_{am}$ for a relative position $x_0=3a/8=1/4$. The "anti-magic" potential takes its mean value half-way between the minima of the "magic" potential (vertical dotted lines). The superposition of these two potentials gives the state-dependent trapping potentials $V_{g,e}= V_m \pm V_{am}$. The positions of the resulting $|g\rangle$-sites and $|e\rangle$-sites are marked by the corresponding blue and red letters. Dash-dotted blue line: triangular unit cell of the state-$|g\rangle$ lattice. Dashed red line: triangular unit cell of the state-$|e\rangle$ lattice. The combination of these two regular triangular lattices produces a honeycomb lattice with alternating state-dependent sites. The unit cell of this honeycomb lattice (black solid line) is no longer regular, see Fig.~\ref{f5}.
}
\end{figure}

Next we require $x_0\!=\!3a/8$ ($=\!1/4$ in dimensionless units) such that $\cos(2\pi x_0)=0$, see Fig. \ref{f3}. From an experimental point of view, this would require to control the phases of the laser beams. For this particular choice of $x_0$, one has (in dimensionless units)
\begin{equation}
V_{am}({\bf r})=V_1 \frac{1+\sin(2\pi x)}{2}.
\end{equation}

The state-dependent potentials satisfy $V_g(x,y)= V_1+V_e(-x,y)$. Since they are also even functions of $y$, they also satisfy $V_g({\bf r})= V_1+V_e(-{\bf r})$. This means that $V_g$ and $V_e$ are thus simply obtained from each other by mere reflection about the $Oy$ axis and also by mere inversion about the origin.

\begin{figure}[!htb]
\centerline{\includegraphics[width=8cm]{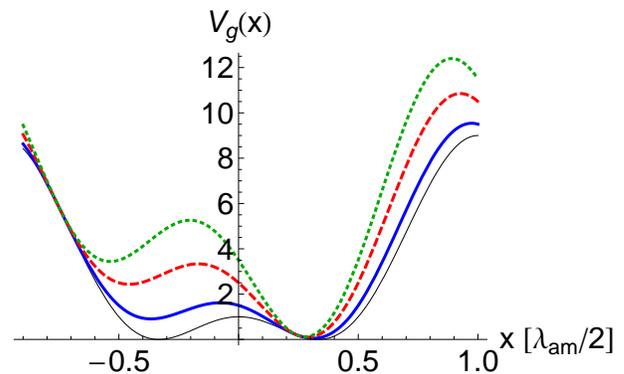}}
\caption{\label{f4} [Color online] Plot of $V_g(x,y)$ as a function of $x$ when $y=1/\sqrt{3}$ for various $V_1/V_0$ ratios. Black thin line: "magic" potential for $V_1=0$. Blue thick line: $V_1=V_0$. Red dashed line: $V_1=3V_0$. Green dotted line: $V_1=5V_0$.
}
\end{figure}

Figure \ref{f4} shows $V_g$ as a function of $x$ when $y=1/\sqrt{3}$ for various $V_1/V_0$ ratios. As $V_1$ is increased from zero, the two wells of the "magic" honeycomb lattice get shifted away and achieve different potential values.
A single band tight-binding description will be appropriate for $V_g$ when the wells are deep enough and sufficiently well separated in energy. The recoil energy associated to the "magic" honeycomb potential is here $E_R = E_m\cos^2\theta$, where $E_m=\hbar^2k^2_m/(2M)$ is the "magic" recoil energy, $M$ being the mass of Yb atoms. Under the current experimental configuration, one has $E_R= 4E_{am}/3$ where $E_{am}=\hbar^2k^2_{am}/(2M)$ is the "anti-magic" recoil energy. Direct tunneling between theses two wells is then suppressed when $V_1\gtrsim V_0 \gg E_{am}$ (we take $V_1=3V_0$ in our subsequent calculations).
One can also show that the harmonic approximation around the global minima of $V_g$ is no longer isotropic but features two different trapping frequencies along $Ox$ and $Oy$. In turn, the corresponding Wannier functions $w_g({\bf r}-{\bf r}_g)$, centered at the $|g\rangle$-sublattice sites ${\bf r}_g$, will reflect this spatial asymmetry. For $V_1=3V_0$, the harmonic anisotropy is small and the corresponding harmonic lengths differ by $10\%$ only. Obviously, the same conclusions apply to $V_e$ and its associated Wannier functions $w_e({\bf r}-{\bf r}_e)$ centered at the $|e\rangle$-sublattice sites ${\bf r}_e$. Furthermore, because of the inversion property between the two optical potentials one can infer the interesting property:
\begin{equation}
\label{InvProp}
w_e(-{\bf r})=w_g({\bf r}).
\end{equation}

\begin{figure}[!htb]
\centerline{\includegraphics[width=6cm]{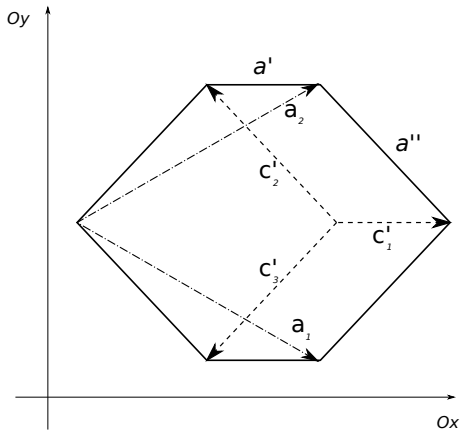}}
\caption{\label{f5} The Wigner-Seitz cell of the new honeycomb structure created by the superposition of the two triangular state-dependent trapping potentials. This unit cell is no longer regular and we have $a'=0.88 a$, $a"= 1.07 a$. The proportions on the figure have been exaggerated.
}
\end{figure}

The superposition of these two independent regular triangular state-$|e\rangle$ and state-$|g\rangle$ sublattices defines a new honeycomb structure and Fig.~\ref{f5} shows its primitive Wigner-Seitz cell. As one can see, contrary to the "magic" potential, the resulting new honeycomb structure is no longer regular as the two sublattices are no longer shifted by $a$ along $Ox$ but by $a' \approx 0.88 a$. The new nearest-neighbor vectors ${\bf c}'_a$ ($a=1,2,3$) no longer add up to zero but, because the underlying Bravais lattice is still triangular, they still verify ${\bf a}_1 = {\bf c}'_1-{\bf c}'_2$ and ${\bf a}_2 = {\bf c}'_1-{\bf c}'_3$. As a consequence, the overlap of the Wannier functions $w_g$ and $w_e$ along the horizontal link will be different from their overlaps along the two other links (these being identical), a fact which has its importance when coupling the two sublattices. 

\section{Uniform flux configuration}
\label{uni}

So far we have been able to produce a honeycomb lattice where each of its sublattice traps atoms of a given internal state. However these sublattices are still uncoupled as atoms cannot yet flip their internal state and hop. For this one needs to expose the atoms to Raman lasers which, by resonantly coupling the two internal states of the atoms, will induce these hops and thus couple the two sublattices. As will be explained below and in Section~\ref{stg}, uniform or staggered synthetic magnetic fields can then be implemented for a suitable choice of the Raman lasers, and atoms traveling around a unit cell will pick up a nonzero phase. The net flux per cell can be made of the order of one quantum flux, thereby reaching the strong field regime.

\subsection{Raman-induced hopping}

The Raman coupling between the two honeycomb sublattices makes an atom with internal state $|g\rangle$ at site ${\bf r}_g$ hop to a site at ${\bf r}_e$ while flipping its internal state to $|e\rangle$ (and vice versa).  For a plane-wave Raman laser field with wavevector ${\bf q}$, the associated (complex) hopping amplitude reads:
\begin{equation}
J_{g\to e}= \frac{\hbar \Omega}{2}\int d^2{\bf r} \, w^*_e({\bf r}-{\bf r}_e) e^{i{\bf q\cdot r}} w_g({\bf r}-{\bf r}_g),
\end{equation}
where $\Omega$ is the Raman laser Rabi frequency. For the reverse hopping process, one has simply $J_{e\to g}= J^*_{g\to e}$. Most importantly, if one now considers the hopping amplitude associated to Bravais-translated sites ${\bf r}'_e =  {\bf r}_e+{\bf R}$ and ${\bf r}'_g =  {\bf r}_g+{\bf R}$, where ${\bf R}$ is a Bravais lattice vector, then:
\begin{equation}
J_{g'\to e'}=  e^{i{\bf q}\cdot {\bf R}} \, J_{g\to e} .
\end{equation}
Another interesting property can be found by defining ${\bf R}_{ge}= ({\bf r}_e+{\bf r}_g)/2$ and ${\bf r}_{ge} = {\bf r}_e-{\bf r}_g$. Then:
\begin{equation}
\label{Jeg}
J_{g\to e}= \frac{\hbar \Omega}{2} e^{i{\bf q\cdot R}_{ge}} \, I_{ge},
\end{equation}
with
\begin{equation}
\label{Ieg}
I_{ge}=\int d^2{\bf r} \, w^*_e({\bf r}+\frac{{\bf r}_{ge}}{2}) e^{i{\bf q\cdot r}} w_g({\bf r}-\frac{{\bf r}_{ge}}{2}).
\end{equation}
Using now Eq.~\eqref{InvProp}, it is easy to show that $ I_{ge}= I^*_{ge}$ meaning that $ I_{ge}$ is in fact real. Assuming the Raman laser Rabi frequency to be real, the phase of the hopping amplitude is then simply given by:
\begin{equation}
\Phi_{g \to e} = - \Phi_{e \to g} = {\bf q} \cdot \frac{{\bf r}_e+{\bf r}_g}{2}.
\end{equation}

In the following, we will assume that the overlap between the Wannier functions is only significant for nearest-neighbor sites. This will be the case when $V_1\gtrsim  V_0\gg E_{am}$. As stated earlier, this overlap, and thus $I_{ge}$, will be link-dependent as the new honeycomb lattice is no longer regular. We will thus neglect direct tunneling or second-order Raman-induced tunneling within each sublattice and only consider nearest-neighbor hopping between the two sublattices.

\begin{figure}[!htb]
\centerline{\includegraphics[width=7cm]{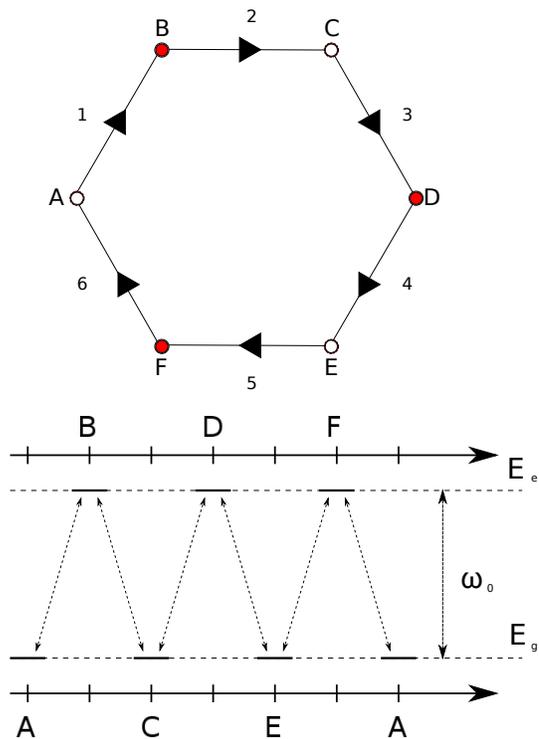}}
\caption{\label{f6} [Color online] Top figure: sketch of the hexagonal plaquette alternating sites where state-$|g\rangle$ atoms are trapped (A, C and E, open circles) and state-$|e\rangle$ atoms are trapped (B, D and F, red-filled circles). The total phase accumulated per cell is calculated for atoms hopping clockwise (arrows). Bottom figure: Energy diagram along the loop and the corresponding Raman transitions. All Raman transitions occur here at the angular frequency separation $\omega_0$ between the two hyperfine state. As a consequence one single Raman laser beam is enough to address all sites.
}
\end{figure}

\subsection{One single Raman laser field is not enough}

Unfortunately, the simplest scheme where one uses a single Raman laser beam {\it does not} induce any global phase around a Wigner-Seitz plaquette. Indeed, considering the situation depicted in Fig. \ref{f6}, one can identify pairs of hopping amplitudes related by a Bravais translation:
\begin{subequations}
\begin{align}
&J_{A\to B} = e^{-i {\bf q}\cdot {\bf a}_1} \, J^*_{D\to E} , \\
&J_{B\to C} = e^{i {\bf q}\cdot ({\bf a}_1-{\bf a}_2)} \, J^*_{E\to F}, \\
&J_{C\to D} = e^{i {\bf q}\cdot {\bf a}_2} \, J^*_{F\to A} .
\end{align}
\end{subequations}
It is then easily seen that the total phase accumulated around the Wigner-Seitz plaquette trivially cancels out and thus this Raman scheme alone fails to produce an artificial magnetic field.

\subsection{More is different}

\begin{figure}[!htb]
\centerline{\includegraphics[width=9cm]{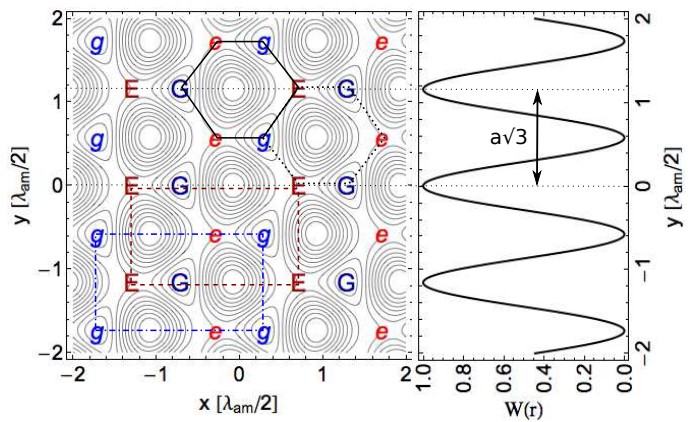}}
\caption{\label{f7} [Color online] The left part of the figure shows the contour plot of the trapping potential $V_g({\bf r})$. The right part of the figure shows the weak perturbing standing-wave potential which is added along $Oy$ to lift the energy degeneracy between the potential wells. The standing-wave potential is maximum for $y=0$ and its period exactly matches $\sqrt{3} a$ where $a$ is the "magic" honeycomb lattice constant. Every well in every other horizontal row is lifted up. Dark blue letters "G " mark the wells which are lifted up whereas blue letters "g" mark the unaffected ones. Accordingly, the dark red letters "E" and red letters "e" mark the respective positions of the lifted-up and unaffected wells for the trapping potential $V_e({\bf r})$. Both state-$|g\rangle$ and state-$|e\rangle$ lattices are rectangular with a two-point unit cell (blue dashed-dotted and red dashed rectangles respectively), the corresponding Bravais vectors being $({\bf a}_1+{\bf a}_2)$ and $({\bf a}_2-{\bf a}_1)$. The resulting global honeycomb structure exhibits two different vertical alternating strips made either of the clockwise "gEGeGE" cell (dotted hexagon) or the "GegEge" one (solid hexagon).
}
\end{figure}

The reason why the single Raman beam scheme fails is because all links are on an equal footing. To cure this problem, we need to consider a slightly more involved Raman laser scheme.
To this end we introduce two counter-propagating laser beams with wavelength $\lambda$ producing a standing-wave pattern $W({\bf r})$ having the periodicity of $V_g$ and $V_e$ along $Oy$. This imposes $\lambda = 2\sqrt{3}a = 2\lambda_{am}/\sqrt{3} \approx 1.29 \mu$m. These laser fields generate different optical potentials for state $|g\rangle$ and $|e\rangle$ (in units of $\lambda_{am}/2$):
\begin{equation}
W_{g,e}({\bf r})=\hbar\Delta_{g,e} \, W({\bf r}) = \hbar\Delta_{g,e} \, \frac{1+ \cos(\pi\sqrt{3}y) }{2},
\end{equation}
where we have asumed that the phases of these additional laser beams are fixed in such a way that the maxima of $W_{g,e}$ along $Oy$ coincide with those of $V_m$ when $x=0$. The new total potentials read:
\begin{equation}
V'_{g,e}({\bf r})= V_m({\bf r})\pm V_{am}({\bf r})+W_{g,e}({\bf r}).
\end{equation}
Because of the chosen periodicity along $Oy$ and choice of phase, the potential value is lifted in every other horizontal row of sites, the potential energy increase being $\hbar\Delta_g$ or $\hbar\Delta_e$ depending on the trapped internal state, see Fig. \ref{f7}.

\begin{figure}[!htb]
\centerline{\includegraphics[width=7.5cm]{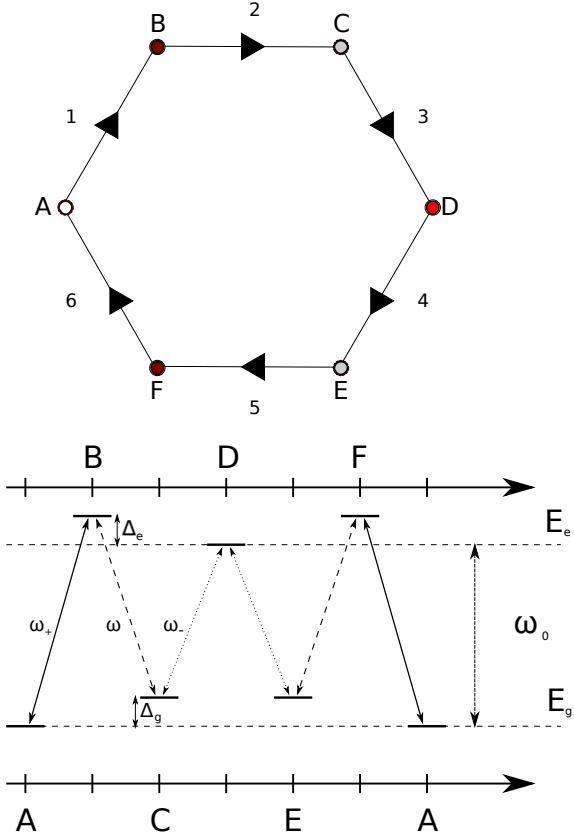}}
\caption{\label{f8} [Color online] Top figure: clockwise "gEGeGE" plaquette. Open and grey-filled circles mark the unaffected and lifted-up $|g\rangle$-sites whereas red-filled and dark-red-filled circles mark the unaffected and lifted-up $|e\rangle$-sites.
Bottom figure: Energy diagram for the Raman transitions around the plaquette. Because the wells are lifted differently, three different angular frequencies, $\omega$, $\omega_+$ and $\omega_-$, are now needed to activate the hops around this plaquette.
}
\end{figure}

\begin{figure}[!htb]
\centerline{\includegraphics[width=7.5cm]{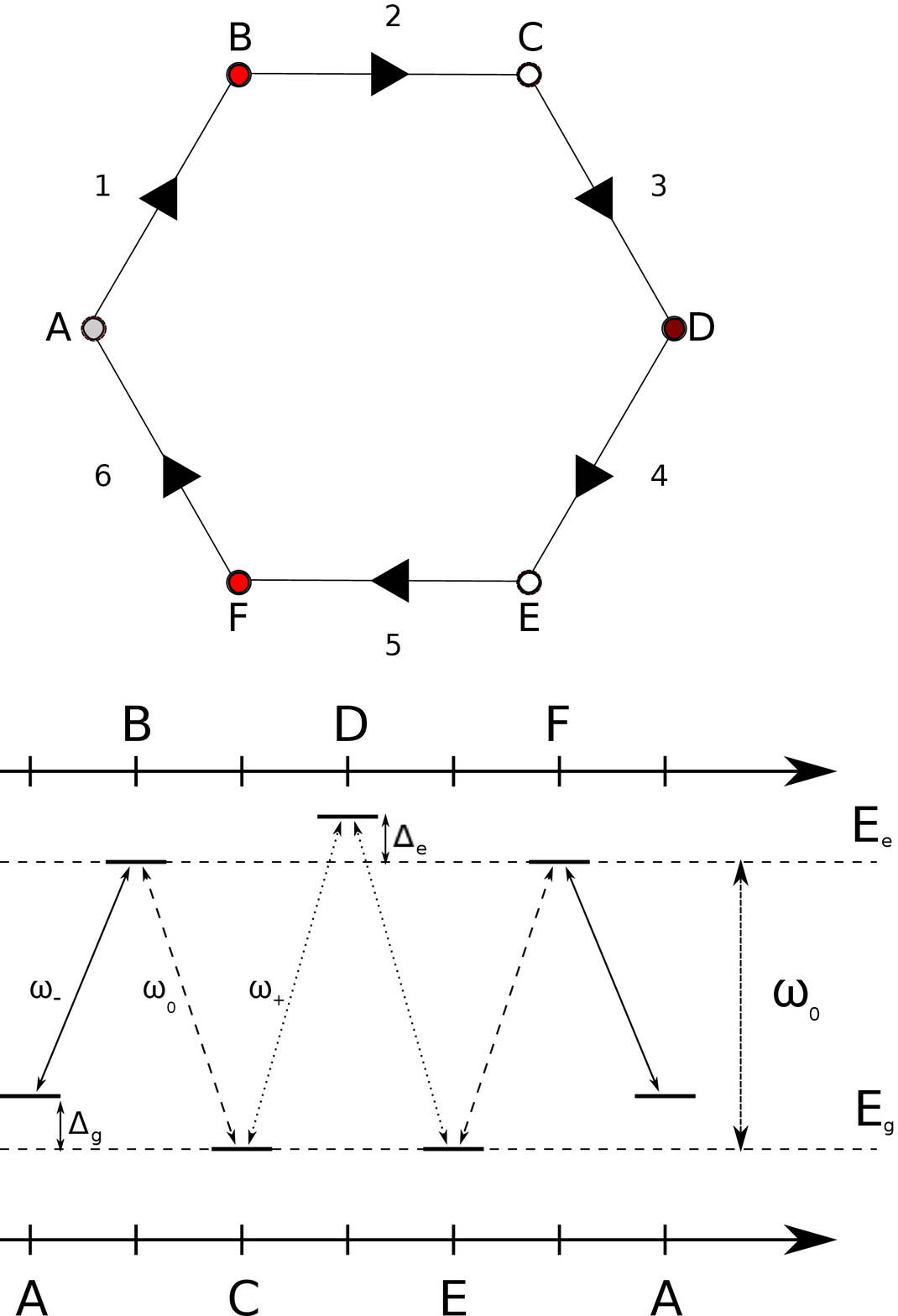}}
\caption{\label{f9} [Color online] Top figure: clockwise "GegEge" plaquette. Open and grey-filled circles mark the unaffected and lifted-up $|g\rangle$-sites whereas red-filled and dark-red-filled circles mark the unaffected and lifted-up $|e\rangle$-sites.
Bottom figure: Energy diagram for the Raman transitions around the plaquette. Because the wells are lifted differently, three different angular frequencies, $\omega_0$, $\omega_+$ and $\omega_-$, are now needed to activate the hops around this plaquette.
}
\end{figure}

In the following we will assume that $\hbar\Delta_{g,e} \ll V_0$ so that the net (perturbative) effect of $W_{g,e}({\bf r})$ is simply to lift the potential energy without modifying the original Wannier functions $w_{e,g}({\bf r})$. However, because of these energy shifts, the honeycomb structure now features two different cells and exhibits vertical "strips" of these cells alternating along $Ox$, see Fig.~\ref{f7}. One now needs four Raman beams to address all lattice links and activate hopping between all neighboring sites, see Fig.~\ref{f8} for one Wigner-Seitz plaquette and Fig.~\ref{f9} for its neighboring one. Their respective angular frequencies are $\omega_+ = \omega_0+\Delta_e$, $\omega_- = \omega_0-\Delta_g$, $\omega = \omega_0+\Delta_e-\Delta_g$ and $\omega_0$ and their respective wavevectors are ${\bf q}_+$, ${\bf q}_-$, ${\bf q}$ and ${\bf q}_0$. For sake of simplicity, we will assume that all these Raman beams propagate in the $(Ox,Oy)$ plane. Since the light-shifts are small compared to the transition angular frequency, $\Delta_{g,e} \ll \omega_0$, it is legitimate to neglect the tiny changes in wavevector length so that we can consider that all these wavevectors have the same norm $k_0=2\pi/\lambda_0$. 

Then, for the Wigner-Seitz cell of Fig.~\ref{f8}, the cumulated total clockwise phase is simply:
\begin{equation}
\Phi = ({\bf a}_2-{\bf a_1})\cdot \frac{{\bf q}_+ +{\bf q}_- -2{\bf q}}{2},
\end{equation}
while the one for that of Fig.~\ref{f9} is:
\begin{equation}
\Phi'= ({\bf a}_2-{\bf a_1})\cdot \frac{{\bf q}_+ +{\bf q}_- - 2{\bf q}_0}{2}.
\end{equation}
Of course we recover the fact that the phases cancel as they should when all wave vectors are identical.
The induced artificial magnetic flux will be uniform provided $\Phi=\Phi'$, i. e. provided $({\bf q}-{\bf q}_0)\cdot ({\bf a}_2-{\bf a}_1)=0$. This will be the case if ${\bf q}_0$ and ${\bf q}$ are most simply chosen parallel. If we now further choose ${\bf q}_+$ and ${\bf q}_-$ to be parallel and anti-parallel to ${\bf q}_0$ and ${\bf q}$, the total cumulated phase then reads $\Phi= 2\pi\alpha$ with:
\begin{equation}
\label{alpha}
\alpha = \frac{2\lambda_{am}}{\sqrt{3}\lambda_0} \, \cos\theta_R.
\end{equation}
where $\theta_R$ is the angle of ${\bf q}_+$ with respect to axis $Oy$. The maximum value for $\alpha$ is $2\lambda_{am}/(\sqrt{3}\lambda_0) \approx 2.24$, meaning that this scheme can provide a bit more than two flux quantum per cell. For comparison the corresponding magnetic field giving rise to the same maximum flux would be $B=8\pi\hbar/(e\lambda_0\lambda_{am}) \approx 256$G. To obtain the same flux in graphene samples, as the graphene lattice constant is very small (of the order of $0.15$nm), one would need a magnetic field about $6\times10^6$ larger.

\section{Staggered flux configuration}
\label{stg}

\begin{figure}[!htb]
\centerline{\includegraphics[width=7.5cm]{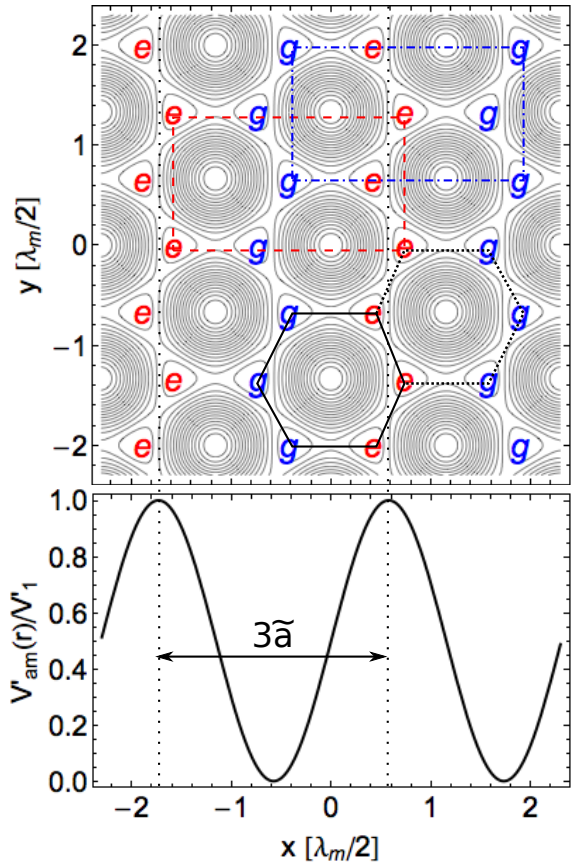}}
\caption{\label{f10} [Color online] Staggered magnetic field case. Contour plot of the "magic" honeycomb lattice $V'_m({\bf r})$ and the "anti-magic" standing-wave potential $V'_{am}({\bf r})$ with spatial period $3\tilde{a}$, $\tilde{a}$ being the "magic" honeycomb lattice constant. The blue letters "g" mark the positions of the state-$|g\rangle$ sites for the trapping potential $V'_g=V'_m+V'_{am}$ while red letters "e" mark the positions of the state-$|e\rangle$ sites for the trapping potential  $V'_e=V'_m-V'_{am}$. Both state-$|g\rangle$ and state-$|e\rangle$ lattices are rectangular but with a two-point unit cell (blue dashed-dotted and red dashed rectangles respectively), the corresponding Bravais vectors being $({\bf a}_1+{\bf a}_2)$ and $({\bf a}_2-{\bf a}_1)$. The Raman laser beams only connect horizontal nearest-neighbor state-$|g\rangle$ and state-$|e\rangle$ sites. The resulting honeycomb structure exhibits two different alternating vertical strips made either of the clockwise cell "eeggge" (dotted hexagon) or the "ggeeg" one (solid hexagon).
}
\end{figure}

\begin{figure}[!htb]
\centerline{\includegraphics[width=8cm]{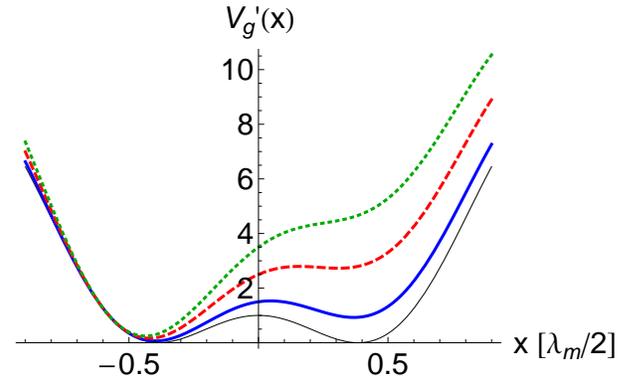}}
\caption{\label{f11} [Color online] Staggered magnetic field case. Plot of the trapping potential $V'_g({\bf r})$ as a function of $x$ when $y=\sqrt{3}\tilde{a}/2$ for various $V'_1/V'_0$ ratios. Black thin line: "magic" honeycomb potential ($V'_1=0$). Blue thick line: $V'_1=V'_0$. Red dashed line: $V'_1=3V'_0$. Green dotted line: $V'_1=5V'_0$.
}
\end{figure}

\begin{figure}[!htb]
\centerline{\includegraphics[width=7cm]{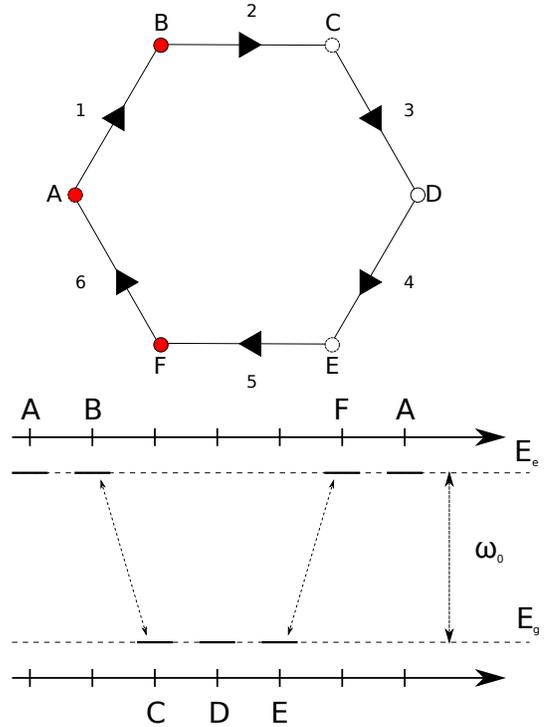}}
\caption{\label{f12} [Color online] Staggered magnetic field case. Top figure: clockwise "eeggge" plaquette. The red-filled circles mark the state-$|e\rangle$ sites while open circles mark the state-$|g\rangle$ sites. Bottom figure: Energy diagram for the Raman transitions around the plaquette. A single Raman laser beam at frequency $\omega_0$ is required here to activate hopping along the horizontal "e-g" links. Hopping along the vertical direction is due to direct tunneling through a potential barrier.
}
\end{figure}

\subsection{Adapting the previous configuration}

An artificial staggered magnetic flux can be easily created using the results of the previous Section. Indeed, by simply taking the ${\bf q}_+$ and ${\bf q}_-$ wave vectors to be along axis $Ox$ and the ${\bf q}_0$ and ${\bf q}$ wave vectors to be anti-parallel, the accumulated phases in the two different type of Wigner-Seitz cell would now be opposite:
\begin{equation}
\Phi'=-\Phi= ({\bf a}_2-{\bf a_1})\cdot {\bf q}= \pi\alpha,
\end{equation}
where $\theta_R$ in Eq.~\eqref{alpha} is now the angle of ${\bf q}$ with respect to $Oy$. We get alternating vertical stripes with opposite fluxes, the maximum value reached being half the uniform one, i. e. about $1.12$ flux quanta. Hence, the configuration proposed in Section~\ref{uni} proves quite versatile as it allows to easily switch from the uniform to the staggered magnetic flux cases by only changing the direction of propagation of the Raman beams.

\subsection{A novel configuration}

However, for sake of completeness, we would like to propose here another scheme, similar to the one proposed in \cite{dalibard}, where the final lattice configuration alternate zig-zag vertical rows of $|g\rangle$-state minima with zig-zag vertical rows of $|e\rangle$-state minima. To achieve this, we simply start with a "magic" honeycomb potential where all beams are now coplanar with respective consecutive angles of $120^\circ$ like done in \cite{keanloon}. With respect to the previous section, this configuration is obtained by setting $\theta=0$, see Eq.~\eqref{vectorsk}. Using now $\lambda_m/2$ as the space unit in this Section, we get $V'_m({\bf r}) = V'_0 [3+2f'_m({\bf r})]$ with:
\begin{equation}
f'_m({\bf r})=2\cos{(\frac{\sqrt{3}\pi x}{2})} \, \cos{(\frac{3\pi y}{2})}+\cos{(\sqrt{3}\pi x}).
\end{equation}
This "magic" honeycomb lattice constant is $\tilde{a}=4/(3\sqrt{3})$ in dimensionless units ($\tilde{a}\approx 0.29 \mu$m in full units).

The "anti-magic" potential is now produced with two laser beams counter-propagating along $Ox$, their off-plane elevation angle being $\theta'$. In dimensionless units, the "anti-magic" potential reads
\begin{equation}
V'_{am}({\bf r})=V'_1 \frac{1+\cos{({\bf b'}\cdot({\bf r}-{\bf r}'_0))}}{2},
\end{equation}
where ${\bf b'}=b' {\bf \hat{e}}_x = 4\pi\cos\theta' /\lambda_{am} \bf \hat{e}_x$. We now request the period of the"anti-magic" standing-wave to match $3\tilde{a}$ and the horizontal shift to be $x'_0=3\tilde{a}/4$. This imposes the off-plane angle to satisfy
\begin{equation}
\cos\theta' = \frac{\sqrt{3} \lambda_{am}}{4\lambda_m},
\end{equation}
giving $\theta' \approx 50.35^\circ$. One then finds:
\begin{equation}
V'_{am}({\bf r})=V'_1 \frac{1+\sin{(\sqrt{3}\pi x/2)}}{2},
\end{equation}
and $V'_{g,e}({\bf r}) = V'_m({\bf r}) \pm V'_{am}({\bf r})$. With this set of parameters, one finds that ${\bf b}' = ({\bf b}_1+{\bf b}_2)/2$. In other words the Bravais lattice associated to the potential $V_g$ and $V_e$ is no longer triangular. It turns out to be rectangular with reciprocal lattice vectors ${\bf b}'$ along $Ox$ and ${\bf b}''= \sqrt{3}b'\bf \hat{e}_y$ along $Oy$ respectively, the Wigner-Seitz cell of the lattice of global minima being a two-point cell, see Fig.~\ref{f10}. As $V_g$ and $V_e$ are still related by inversion and reflection about $Oy$, the inversion symmetry relating their associated Wannier functions remains valid. Figure \ref{f11} shows a plot of $V'_g$ when $y=\sqrt{3}\tilde{a}/2$ for various ratios $V'_1/V'_0$. A single band description for each rectangular lattice is appropriate when the condition $V'_1 > V'_0 \gg E_m$ is met, $E_m$ being the "magic" recoil energy. Typically a ratio $V'_1/V'_0 =3$ or larger is required. 

To couple the two shifted independent rectangular $|g\rangle$-state and $|e\rangle$-state sublattices, it is then sufficient to use a single in-plane Raman laser at frequency $\omega_0$ and wave vector ${\bf k}_0$. We assume that the Raman laser only couples Yb atoms (with different internal states) which are trapped in nearest-neighbor sites. As Fig.~\ref{f10} shows, the Raman-induced hops only occurs along $Ox$, between vertical zig-zag rows. The hopping along the zig-zag rows relies on direct tunneling through the potential barriers. As a whole, the coupled system displays a honeycomb structure. Each vertical row is built by repeated tiling of the same plaquette. There are two different kind of plaquettes, obtained from each other by reflection about their middle vertical axis. The rows with different plaquettes are alternating along $Ox$. A schematic picture of the hops around the clockwise "eeggge" plaquette is shown in Fig. \ref{f12}. The situation for the other "ggeeeg" plaquette is simply obtained from it by mirroring the sites through the middle vertical axis and changing the energy diagram accordingly.

By an argument similar to the case of the uniform synthetic magnetic flux, the phase of the hopping amplitude between a $|g\rangle$-state site and its neighboring $|e\rangle$-state site is simply given by ${\bf k}_0\cdot {\bf R}_{ge}$, where is the middle point of the (horizontal) connecting link, the phase for the reverse hopping process being the opposite. As the hopping amplitude along the vertical zig-zag chains is real, the total phase accumulated around the cell shown in Fig.~\ref{f10} is simply $\Psi = 2\pi\alpha' = (4\pi\lambda_{m}/3\lambda_0) \cos\theta'_R$, where $\theta'_R$ is the angle of the Raman wave vector with axis $Oy$. The phase for the other plaquette is $-\Psi$. The maximum value obtained for $\alpha'$ with this scheme is $(2\lambda_{m}/3\lambda_0) \approx 0.876$, a bit less than a flux quantum. Here again we get alternating vertical stripes with opposite fluxes.

\section{Harper model}
\label{harpermodel}

In this section we would like to give the single-band Harper model \cite{harper, hofstadter, rammal} describing the previous configurations and discuss some orders of magnitude and limitations. We will restrict ourselves to the configuration obtained in Section~\ref{uni} as it can as well describe a uniform or a staggered synthetic magnetic flux applied to a honeycomb lattice.

To find the Harper Hamiltonian associated to the considered optical potential configuration, one has first to identify its unit cell $C$ and the corresponding Bravais lattice $\mathcal{B}$. Then the Harper Hamiltonian is simply written as:
\begin{equation}
\label{Hharper}
H_{\mathrm{Harper}} = \sum_{n\in \mathcal{B}} \, H_n,
\end{equation}
where $H_n$ is the hop operator acting on the cell $C_n$ obtained by translation of $C$ along $\mathcal{B}$. In turn, one can write:
\begin{equation}
\label{hop}
H_n = T_n + T^\dag_n,
\end{equation}
where $T_n$ describes hops among sites within $C_n$ but along prescribed directions.

\subsection{Simplest uniform flux configuration}

\begin{figure}[!htb]
\centerline{\includegraphics[width=4cm]{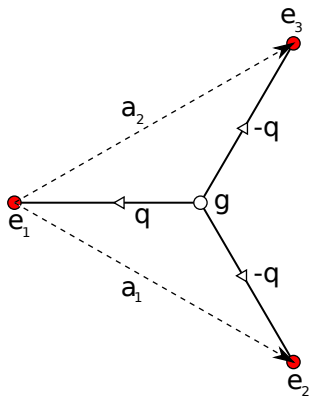}}
\caption{\label{f13} [Color online] The cell used to derive the Harper Hamiltonian for the simplest uniform magnetic flux configuration. Red circles: state-$|e\rangle$ sites. Open circle: state-$|g\rangle$ site. The arrows on the links indicate the hop directions considered to write down the hop operator $T_n$, see Eq.~\eqref{hop}. The Raman wave vectors implied in the hops are indicated along each link. The hop operator $T^\dag_n$ operates on the same cell but with reversed hop directions and thus complex conjugate hopping amplitudes. The Harper Hamiltonian is obtained by superposing the hop operators obtained by repeated translations along the Bravais lattice $\mathcal{B}$ spanned by vectors ${\bf a}_1$ and ${\bf a}_2$, see Eq.~\eqref{Hharper}.
}
\end{figure}

This is the configuration obtained for ${\bf q} = {\bf q}_0 = -{\bf q}_+ = - {\bf q}_-$. In this case, it is easy to see that the lattice is obtained by repeated tiling of the unit cell displayed in Fig.~\ref{f13}, the relevant Bravais lattice $\mathcal{B}$ being triangular and spanned by ${\bf a}_1$ and ${\bf a}_2$. We take the origin of the cell at the state-$|g\rangle$ site located at ${\bf r_n}$. We denote by $g_n$ the corresponding annihilation operator and by $e_{an}$ the annihilation operators of a state-$|e\rangle$ atom located at sites ${\bf r}_{an}={\bf r_n}-{\bf c}'_a$ ($a=1,2,3$). The hop operator is:

\begin{equation}
\label{UniformHop}
T_n = \sum_a J_a e^{i\phi_{an}} \, e^\dag_{an}g_n,
\end{equation}
where $J_a$ is the (real) hopping strength along the link vector $-{\bf c}'_a$ and $\phi_{an} = \varepsilon_a \, {\bf q}\cdot({\bf r}_n - {\bf c}'_a/2)$ ($\varepsilon_2=\varepsilon_3 = -\varepsilon_1= 1$). As obtained from Section~\ref{uni}, the total anti-clockwise phase per unit cell is:

\begin{equation}
\Phi_B = 2{\bf q}\cdot({\bf a}_2-{\bf a}_1).
\end{equation}
At this point one may wonder which gauge potential ${\bf A}$ and which magnetic field ${\bf B}= {\boldsymbol \nabla} \times {\bf A}$ would give rise to the same Harper model for electrons in the graphene lattice. Using the gauge potential ${\bf A}_L= -By\hat{{\bf e}}_x$ (Landau gauge), we write ${\bf A} = {\bf A}_L+{\bf A}_0$. The vector potential ${\bf A}_L$ generates a uniform magnetic field perpendicular to the honeycomb lattice and we choose its strength $B$ such that it gives rise to the same flux per unit cell as ${\bf A}$. It is then easy to show that the vector potential ${\bf A}_0$, which does give rise to a nonzero magnetic field, nevertheless gives rise to zero flux around any closed loop. In the Harper model the particles move along the lattice links, it means that the contribution of ${\bf A}_0$ can in fact be locally gauged away. Indeed, picking up some lattice site ${\bf r}_0$ as the origin, the phase $\Phi_n = \int_{{\bf r}_0}^{{\bf r}_n}{\bf A}_0\cdot d{\bf r}$ is in fact path independent (as long as the path is taken on the honeycomb lattice) and only depend on the end point ${\bf r}_n$. This proves that the Harper model given by Eqs \eqref{Hharper}, \eqref{hop} and \eqref{UniformHop} can be in fact obtained with a uniform magnetic field.

We would like to give now a somewhat simpler expression for the hop operator. Performing the local $U(1)$ gauge transformation:
\begin{equation}
\begin{split}
&g_n \to \exp(-i {\bf q}\cdot {\bf r}_n/2) \, g_n\\
&e^\dag_{an} \to \exp(-i {\bf q}\cdot {\bf r}_{an}/2) \, e^\dag_{an},
\end{split}
\end{equation}
one gets:
\begin{equation}
T_n = J_1 e^{i\varphi} \, e^\dag_{1n}g_n  + J_2 \, e^\dag_{2n}g_n + J_3  \, e^\dag_{3n}g_n
\end{equation}
where $\varphi = {\bf q}\cdot ({\bf r}_n+{\bf r}_{1n})$ is the Raman phase evaluated at the mid-point along the link vector $-{\bf c}'_1$. One can easily check that the total anti-clockwise phase picked up around a hexagonal plaquette is again $\Phi_B$. This local $U(1)$ gauge transformation amounts to choose locally the phase of the Wannier functions.

Because the link vectors ${\bf c}'_2$ and  ${\bf c}'_3$ have same length, the tunneling amplitudes $J_2$ and $J_3$ are the same as long as the corresponding Raman Rabi frequencies are equal. However, as  ${\bf c}'_1$ has a different magnitude, in principle $J_1$ is different from the two others. The mismatch could be compensated for by a fine-tuning of the corresponding Raman Rabi frequency though this might not be easy in practice. In another context, it is known that hopping imbalance can have a strong impact on the physical phenomena under study \cite{dietl}.

Forgetting about this hopping imbalance, one can give a rough order of magnitude of the hopping amplitude $J = ( \hbar\Omega/2) I_{eg}$, see Eqs.~\eqref{Jeg} and \eqref{Ieg}. Using a harmonic approximation, and further neglecting the Wannier functions anisotropy which is anyway small, a simple calculation performed at $V_1=3V_0$ shows that:
\begin{equation}
J \approx \frac{\hbar\Omega}{2}\, \exp[-(1.587/\hbar_e+0.323 \hbar_e)],
\end{equation}
where the effective Planck's constant is $\hbar_e = \sqrt{2E_{am}/V_0}$ and $E_{am}$ is the anti-magic recoil energy ($E_{am}/h = 900$Hz). For $\hbar\Omega = 2E_{am}$ and $V_0=16E_{am}$, one gets $J\approx 10^{-2}E_{am}$.

\subsection{General case}

\begin{figure}[!htb]
\centerline{\includegraphics[width=7cm]{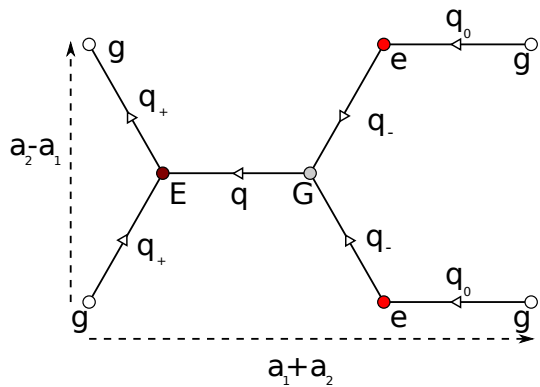}}
\caption{\label{f14} [Color online] The cell used to to derive the Harper Hamiltonian in the general case. Red circles: state-$|e\rangle$ sites. Dark red circles: lifted-up state-$|e\rangle$ sites. Open circles: state-$|g\rangle$ sites. Grey circles: lifted-up state-$|g\rangle$ sites. The arrows on the links indicate the hop directions considered to write down the hop operator $T_n$, see Eq.~\eqref{hop}. The Raman wave vectors implied in the hops are indicated along each link. The hop operator $T^\dag_n$ operates on the same cell but with reversed hop directions and thus complex conjugate hopping amplitudes. The Harper Hamiltonian is obtained by superposing the hop operators obtained by repeated translations along the rectangular Bravais lattice $\mathcal{B}$ spanned by vectors ${\bf a}_1+{\bf a}_2$ and ${\bf a}_2-{\bf a}_1$, see Eq.~\eqref{Hharper}. 
}
\end{figure}

In the general case where all Raman wave vectors are different, the unit cell must now be enlarged and it is given in Fig.~\ref{f14}. Writing down the hop operator $T_n$ is straightforward but rather tedious as now many links are required. We leave it as an exercise for the reader. The noticeable point is that the relevant Bravais lattice $\mathcal{B}$ is now rectangular and spanned by the vectors ${\bf a}_1+{\bf a}_2$ and ${\bf a}_2-{\bf a}_1$.

The above Harper model can be readily extended when interactions come into play. This could be experimentally realized by loading the lattice with bosonic atoms in the presence of on-site repulsive interactions. In this case one could target the fractional quantum Hall effect and the Laughlin state at filling fraction $\nu = 1/3$, and more generally highly-correlated quantum liquids. By loading the lattice with fermionic atoms, one could target the relativistic quantum Hall effect as evidenced in graphene samples at low fluxes per cell. It is worth mentioning that contrary to $^{174}$Yb which has zero total spin, the fermionic Yb isotopes have a nuclear spin. They are thus multi-level systems and one can then think of designing more elaborate configurations to mimic non-Abelian gauge fields. One could also imagine loading the lattice with Bose-Fermi mixtures like $^{173}$Yb$-$$^{174}$Yb~\cite{fukuhara1} or like $^{171}$Yb$-$$^{174}$Yb where next-nearest-neighbor interactions of the order of six times the tunneling energy have been reported at zero magnetic field~\cite{kitagawa}. The case of Fermi-Fermi mixtures raise tantalazing questions on spinor superfluidity. For instance, the $^{171}$Yb$-$$^{173}$Yb mixture realizes a system with $SU(2)\times SU(6)$ symmetry \cite{fukuhara, taie}.

\section{Conclusion}
\label{sum}

In this paper we have proposed experimental set-ups realizing Abelian gauge fields acting on Yb atoms moving in a honeycomb lattice and giving rise to uniform or staggered synthetic magnetic fluxes. A net flux per unit cell of one quantum flux can be easily reached, thereby realizing the cold atom analogue of the Harper model. Different phenomena could be experimentally studied with these configurations, ranging from the relativistic to the fractional quantum Hall effects. A possible extension of this work would be to study the role of the honeycomb lattice distortion which lead to hopping strength imbalance. As far as we know the impact of this imbalance on the Harper model is largely unexplored but would be of great experimental relevance. 

\acknowledgments

ChM wishes to thank Mark Goerbig (LPS, France) for a helpful overseas discussion about quantum Hall effects and St\'ephane Bressan (SoC, NUS) for his interest in the work. BG and ChM acknowledge support from the CNRS PICS Grant No. 4159 and from the France-Singapore Merlion program, FermiCold grant No. 2.01.09. ChM is a Fellow of the Institute of Advanced Studies at NTU. The Centre for Quantum Technologies is a Research Centre of Excellence funded by the Ministry of Education and the National Research Foundation of Singapore.

\end{document}